\documentclass[prd,showpacs,preprintnumbers,amsmath,amssymb,twocolumn,superscriptaddress,notitlepage,nofootinbib,longbibliography,reprint]{revtex4-2}
\usepackage{graphicx,color}
\usepackage{graphicx}
\usepackage{amsmath,amssymb,amsfonts}
\usepackage{stackrel}
\usepackage{enumitem}
\usepackage[english]{babel}
\usepackage{verbatim}
\usepackage{xcolor}
\usepackage{dcolumn}
\usepackage{physics}
\usepackage{verbatim}
\usepackage{float}
\usepackage{color}
\usepackage{multirow}
\usepackage{graphicx}
\usepackage{epstopdf}
\usepackage{epsfig}
\usepackage{slashed}
\usepackage{physics}
\usepackage{multirow}
\usepackage{makecell}
\usepackage{placeins}
\usepackage{setspace}
\usepackage{tabularx}
\usepackage{colortbl}
\usepackage{float}
\usepackage{placeins}
\usepackage{ragged2e}
\usepackage{listings}
\usepackage{gensymb}
\usepackage{siunitx}
\usepackage{aas_macros}
\usepackage{bm}
\usepackage{topcapt}
\usepackage{fontawesome5}
\usepackage[colorlinks=true, allcolors=blue]{hyperref}

\definecolor{darkred}{rgb}{.7,.1,.1}
\usepackage{hyperref}
\hypersetup{colorlinks=true,
	breaklinks=true,
	pdfstartview=Fit,
	linkcolor=blue,
	citecolor=blue,
	urlcolor=blue}
\usepackage{cleveref}

\definecolor{dark-green}{rgb}{0.1,0.7,0.3}

\newcommand{\beq}{\begin{equation}}
\newcommand{\eeq}{\end{equation}}
\newcommand{\bal}{\begin{align}}
\newcommand{\eal}{\end{align}}
\newcommand{\bit}{\begin{itemize}}
\newcommand{\eit}{\end{itemize}}
\newcommand{\ben}{\begin{enumerate}}
\newcommand{\een}{\end{enumerate}}

\renewcommand{\eqref}[1]{Eq.~(\ref{eq:#1})}

\newcommand{\secref}[1]{Sec.~\ref{sec:#1}}

\newcommand{\appref}[1]{Appendix~\ref{sec:#1}}
\newcommand{\figref}[1]{Fig.~\ref{fig:#1}}

\newcommand{\tabref}[1]{Table~\ref{tab:#1}}

\begin{document}

\preprint{}

\title{Galactic Center gamma-ray excess from a generic triaxial halo}

\author{Leo Qiyuan Hu}
\thanks{Contact author: leo.hu@my.cityu.edu.hk}
\affiliation{Department of Physics, City University of Hong Kong, Kowloon, Hong Kong SAR, China}

\author{Ilias Cholis}
\thanks{Contact author: cholis@oakland.edu}
\affiliation{Department of Physics, Oakland University, Rochester, Michigan 48309, USA}

\author{Yi-Ming Zhong}
\thanks{Contact author: yiming.zhong@cityu.edu.hk}
\affiliation{Department of Physics, City University of Hong Kong, Kowloon, Hong Kong SAR, China}

\date{\today}

\begin{abstract}
Recent studies of Galactic surveys, such as \textit{Gaia}, have revealed that the Milky Way's gravitational potential comes from a matter distribution that is triaxial and rotated with respect to the Galactic center-Sun axis. This, in turn, could mean that the dark matter halo also shares these properties. 
In this work, by fitting to the \textit{Fermi}-LAT gamma-ray observations, we test the compatibility of the morphology of the Galactic Center excess (GCE) from dark matter annihilation with a triaxial dark matter halo. In particular, we consider both untilted triaxial halos and halos whose principal axes are tilted with respect to the Galactic disk. In our fits of the \textit{Fermi}-LAT data, by testing over a large library of galactic diffuse emission models, we quantify how the halo triaxiality and tilt affect the line-of-sight-integrated annihilation signal and, consequently, the preferred GCE spatial templates. We find that the GCE spectrum and inner cuspiness are robust against variations in the triaxiality and tilt of the dark matter halo. However, in terms of its overall morphology, the GCE in the gamma-ray data can discriminate between choices for the dark matter halo's triaxiality and tilt. Finally, we find that the GCE is more compatible with originating from a triaxial and tilted halo of dark matter than originating from a triaxial and tilted halo of stars, a result important for understanding the GCE's origin.
\end{abstract}

\maketitle

\section{Introduction}
\label{sec:Introduction}

An excess of gamma-ray photons in the GeV energy range toward the center of the Milky Way (MW), known as the Galactic Center excess (GCE), has been persistently observed by the \textit{Fermi} Large Area Telescope (\textit{Fermi}-LAT)~\cite{1999APh....11..277G} for over a decade~\cite{Goodenough:2009gk, 2009arXiv0912.3828V, Hooper:2010mq, Abazajian:2010zy, Gordon:2013vta}. Evidence for its existence is robust~\cite{Calore:2014xka, Zhou:2014lva, TheFermi-LAT:2015kwa, DiMauro:2021raz, Cholis:2021rpp}; in fact, the \textit{DAMPE} telescope independently confirmed the presence of the GCE in its observations \cite{Alemanno:2025jij}. However, the origin of the excess remains under debate.
The three explanations that have been proposed are (1) it arises from dark matter annihilation \cite{Hooper:2011ti, Hooper:2013rwa, Gordon:2013vta, Daylan:2014rsa, Calore:2014xka, Calore:2014nla, Agrawal:2014oha, Berlin:2015wwa, TheFermi-LAT:2017vmf, Karwin:2016tsw, Leane:2019xiy}, or at least predominantly from dark matter annihilation \cite{Cholis:2021rpp, Zhong:2024vyi}, which would provide the first evidence of dark matter interacting with ordinary matter beyond gravity, (2) it is emitted by an unresolved population of faint millisecond pulsars (MSPs) at the Galactic Center (GC)~\cite{Abazajian:2012pn, Petrovic:2014xra, Lee:2015fea, Bartels:2017vsx, Murgia:2020dzu, Gautam:2021wqn, Macias:2023qqc}, for which, however, there are counter-arguments based on our understanding of the properties of MSPs from x-ray and other gamma-ray observations \cite{Hooper:2013nhl, Cholis:2014noa, Cholis:2014lta, Zhong:2019ycb, Buschmann:2020adf, Hooper:2021kyp, List:2025qbx}, or (3) it is the result of bursts of cosmic rays possibly associated with the supermassive black hole's environment \cite{Petrovic:2014uda, Carlson:2014cwa, Cholis:2015dea}. 
These explanations can reproduce the observed spectral features of the GCE, but differ in their expectations for the morphology of the GCE, in their predictions of the expected gamma-ray point sources to be detected from the inner galaxy, as well as in possible associated observations in other wavelengths and other probes, such as cosmic rays. 

Methods to differentiate between the MSP hypothesis, the dark matter hypothesis, and the cosmic-burst hypothesis can be broadly classified into three categories. 
The first category focuses on identifying small-scale power in the GCE, where dark matter annihilation and the cosmic-burst hypotheses predict a smoother photon distribution while MSPs predict a clumpier one (see e.g. \cite{Zhong:2019ycb,  Buschmann:2020adf, Ramirez:2024oiw, Christy:2024gsl, List:2025qbx}). The second category involves examining the morphology of the GCE, that is, its angular distribution, which could differentiate among all three hypotheses.  
In the simplest annihilation scenario, the signal traces the dark matter density squared and is therefore expected to be approximately spherically symmetric around the GC, reflecting an approximately spherical halo shape. In contrast, the MSP scenario predicts a more ``boxy" or elongated peanut shape along the disk bulge inherited from the stellar population (e.g.~\cite{Song:2024iup}). Finally, the cosmic-ray bursts would give some preference for an elongation of the GCE perpendicular to the Galactic plane. There are also hybrid methods that combine these first two approaches~\cite{Manconi:2025ogr}. 

Finally, if the GCE is a signal of weakly interacting dark matter with its proper thermal relic cross section (see \cite{Steigman:2012nb}) as its observed flux suggests (see e.g. \cite{DiMauro:2021raz, Cholis:2021rpp}), then associated signals can be searched for in other targets such as dwarf galaxies, the Andromeda galaxy or the isotropic gamma-ray background \cite{McDaniel:2023bju, Karwin:2020tjw, Cholis:2024hmd}, in other wavelengths \cite{Bringmann:2014lpa}, in cosmic rays \cite{Hooper:2014ysa, Cirelli:2014lwa, Bringmann:2014lpa, Cholis:2020twh, Krommydas:2022loe, Zhu:2022tpr}, as there may in fact already be a signal in cosmic-ray antiprotons \cite{Cuoco:2016eej, Cui:2016ppb, Cholis:2019ejx}; and in direct dark matter and collider searches (for recent works see e.g.~\cite{Berlin:2025fwx, Hooper:2025fda, Roux:2025wem, Hu:2025thq, DiMauro:2025jia, Roy:2025zvo}).  
Additionally, observations from future observatories may shed light on this active debate \cite{Keith:2022xbd, Miller:2023qph, Lei:2025jsu, IceCube:2025fcn, SKAOPulsarScienceWorkingGroup:2025syv}.

In this work, we focus on the morphology of the GCE. In recent years, several alternatives to the boxy bulge have been proposed, including a boxy bulge plus nuclear bulge~\cite{Bartels:2017vsx}, F98~\cite{Macias:2016nev}, X-shape~\cite{Macias:2016nev, Macias:2019omb}, and peanut-shape~\cite{Coleman:2019kax}. In Ref.~\cite{Zhong:2024vyi}, using the library of Galactic diffuse emission (GDE) models from Ref.~\cite{Cholis:2021rpp}, we systematically evaluated different bulge profiles under various masks for the gamma-ray point sources. We found that while the best-fit model generally depends on the GDE choice, both the spectrum and the overall spherical morphology of the GCE remain largely robust against different masks (see also \cite{McDermott:2022zmq}).

In this work, we consider a complementary complication for the morphology-based approach, namely the possibility that the MW dark matter halo is triaxial rather than spherical~\cite{Bland-Hawthorn:2016lwg, Allgood:2005eu, Peter:2012jh, 2013MNRAS.431.1143D, 2019MNRAS.484..476C, Muru:2025vpz}. Triaxiality is a generic expectation of hierarchical structure formation, in which anisotropic collapse and assembly history can drive halos away from spherical symmetry~\cite{Sheth:1999su, Sheth:2001dp, Musso:2019zmr}. Past mergers can also induce triaxial morphology~\cite{Lau:2020qox}. Conventionally, the triaxial shape has been associated with the outer MW halo and was not expected to affect the inner region. However, recent cosmological \textit{N-body} simulations demonstrate otherwise~\cite{Hussein:2025xwm}. The inner halo, shaped by gravitational interactions with the baryonic disk and bar as well as feedback effects, can also exhibit significant triaxiality.  
Rotations of a triaxial rigid body can induce instabilities in the direction of its principal axes (e.g., the tennis racket theorem states that rotation about the intermediate axis is unstable~\cite{goldstein1950classical}), potentially resulting in a tilted configuration in which the principal axes do not align with the Galactic plane. 
Recent studies based on Galactic surveys, such as~\emph{Gaia}~\cite{2016A&A...595A...1G}, DESI~\cite{DESI:2019jxc}, and H3~\cite{2019ApJ...883..107C}, have modeled the phase spaces of stellar halos and stellar streams, both of which are sensitive to the central gravitational potential of visible and dark matter. These analyses indicate that the MW dark matter halo is both triaxial and tilted~\cite{ 2022AJ....164..249H,2023MNRAS.521.4936K,Han:2023phi,Nibauer:2025jvo,2025arXiv251000095D, Putney:2025mch,Li:2025oxg}.\footnote{Some studies, e.g.,~\cite{2022AJ....164..249H, Li:2025oxg}, focus on inferring the 3D morphology of the stellar halo. Simulations suggest that the morphologies of the dark matter halo and the stellar halo are closely correlated~\cite{2022ApJ...934...14H,Han:2023phi, 2025arXiv251000095D}.} Some studies further suggest that a tilted and tumbling halo can explain the warp of the Galactic disk~\cite{2009ApJ...703.2068D,Han:2023phi, 2026arXiv260117599J}.

We investigate how dark matter halo triaxiality and tilt can impact the morphological interpretations of the GCE. We consider two scenarios. First, we examine tilted halo models motivated by recent studies of stellar populations and stellar streams. For these tilted cases, we fix the triaxial axis ratios and the orientation angles (known as yaw and pitch) to representative literature values, while allowing the inner slope/cusp to vary. Moreover, we study an untilted triaxial halo with varying ellipticities and cuspiness, aspects previously investigated \cite{Calore:2014xka, Cholis:2021rpp, Zhong:2024vyi}, but which we explore further. 
These tests will help identify the correct dark matter annihilation morphology and assess how robust the GCE observations are to halo tilting. The paper is structured as follows: In Sec.~\ref{sec:setup}, we set up the density profiles for the untilted and tilted halos. In Sec.~\ref{sec:fitting}, we describe the fitting procedure. Section~\ref{sec:results} presents our main results, and we conclude in Sec.~\ref{sec:discussion}.

\section{Gamma-ray emission from a triaxial halo}
\label{sec:setup}
\begin{figure}
    \centering
    \includegraphics[width=0.8\linewidth]{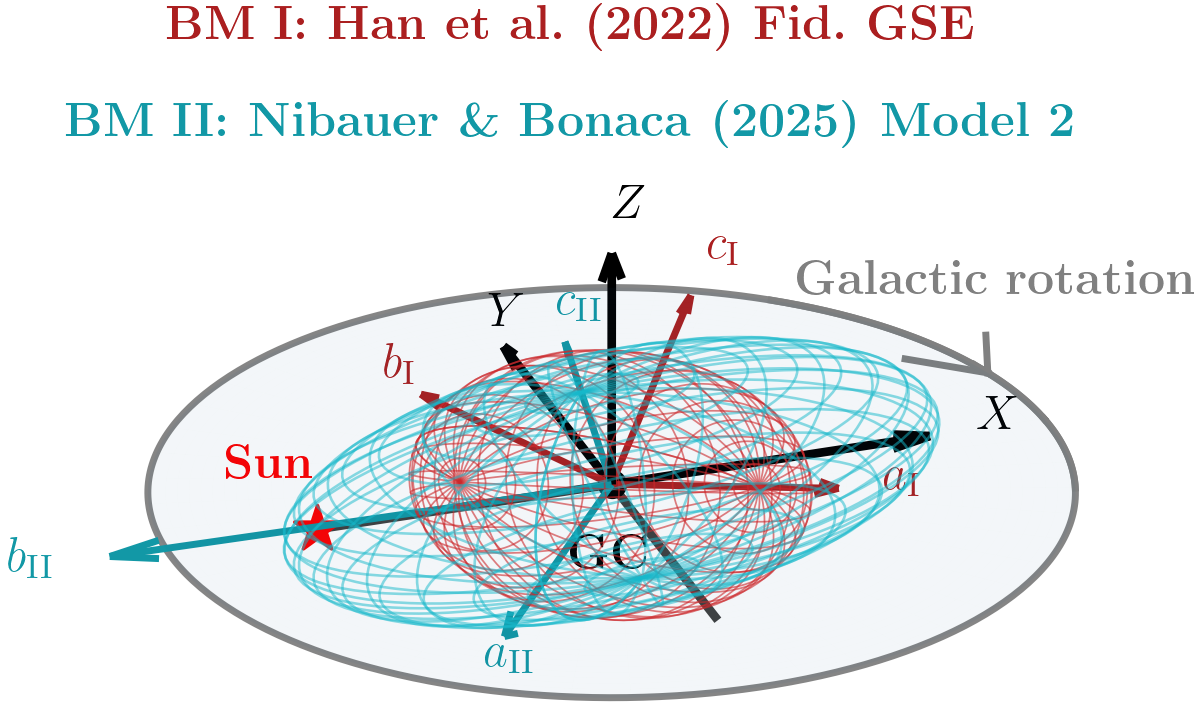}
    \caption{3D view of the tilted triaxial halo of BM I (red) and BM II (cyan). $a_i$, $b_i$, $c_i$ ($i=\text{I}, \text{II}$) represent the directions of major, intermediate, and minor axes of the halos, respectively.}
    \label{fig:benchmark}
\end{figure}

We adopt a galactocentric Cartesian coordinate system to model a generic triaxial halo, with the origin set at the GC. The $X$-axis points from the Sun toward the GC, and the $Z$-axis is perpendicular to the Galactic plane, pointing toward the North Galactic Pole. The $Y$-axis lies in the Galactic plane and is defined such that $(X, Y, Z)$ forms a right-handed coordinate system; it points approximately in the direction of the Galactic rotation, as shown in~\figref{benchmark}. In this setup, the Sun, whose galactocentric distance $r_\odot \equiv \sqrt{x_\odot^2+y_\odot^2+z_\odot^2}=8.5\,\text{kpc}$, is located at $\bm x_\odot = (-8.5\,\text{kpc}, 0 , 0)$. The quadrant $Y>0, Z>0$ corresponds to $\ell > 0, b>0$ in heliocentric Galactic coordinates. 

We model the dark matter halo as an ellipsoid consisting of a series of \emph{concentric} ellipsoidal shells centered at the GC. 
All shells share the same axis ratios $a:b:c$ for the major, intermediate, and minor axes, and the directions of the three principal axes are aligned across all shells. We also assume the halo has reflection symmetries with respect to the mid-planes perpendicular to its principal axes.  We leave more complicated scenarios, such as twisted halos in which the principal axes of each shell are misaligned, to future studies. We introduce the flattening parameters $p\equiv b/a$ and $q\equiv c/a$, which lie in the range $0 <  q \leq p \leq 1$.

Given such a triaxial halo, we model its density profile using a generalized triaxial form based on~\cite{1987gady.book.....B, Lee:2002wb},
\begin{equation}
\rho(r_e) =  \rho_s (r_e/r_s)^{-\gamma} (1+r_e/r_s)^{\gamma-3},
\label{eq:density}
\end{equation}
where $\rho_s$ and $r_s$ are the scale density and radius, respectively, and $\gamma$ is the cuspiness parameter. $r_e$ is the ellipsoidal radius, which is given by,
\begin{equation}
r_e (x,y,z)= \sqrt{\begin{pmatrix}
x & y & z
\end{pmatrix} \bm{R}^\intercal
\begin{pmatrix}
1 & 0&0 \\
0& p^{-2} &0 \\
0& 0& q^{-2}\\
\end{pmatrix} \bm{R} \begin{pmatrix}
x \\
y \\
z \\
\end{pmatrix}}
\label{eq:radius}
\end{equation}
for a general point $(x,y,z)$ in the galactocentric coordinates. Here, $\bm{R}$ is a general 3D rotation matrix representing the tilting of the triaxial halo. In the absence of any tilting ($\bm R = \bm 1$), the unit vectors of major ($\bm a$), intermediate ($\bm b$), and minor axes ($\bm c$) of the triaxial halo align with the unit vectors of the $X$, $Y$, and $Z$ axes, respectively. For such an untilted halo with $p = q = 1$, the generalized triaxial profile of~\eqref{density} reduces to the generalized Navarro–Frenk–White profile~\cite{Navarro:1995iw}.

In the literature, a 3D rotation is commonly decomposed into so-called ``yaw$\to$pitch$\to$roll'' \emph{intrinsic} rotations, where ``intrinsic'' means that the rotations are performed with respect to the (rotated) principal axes of the ellipsoid rather than the fixed extrinsic $XYZ$ coordinate axes. Studies of the stellar halo and stellar streams often find $b\approx c$ and poorly constrain the roll angle. Consequently, the roll is often neglected (more explicitly, the roll is set to $0^\circ$), simplifying the 3D rotation to ``yaw$\to$pitch.'' The corresponding rotation matrix is,
\begin{equation}
\bm{R} = \begin{pmatrix}
\cos \theta & 0 & -\sin \theta \\
0 & 1 & 0 \\
\sin \theta & 0 & \cos \theta\\
\end{pmatrix} 
\begin{pmatrix}
\cos \phi & \sin \phi & 0 \\
-\sin \phi & \cos \phi & 0 \\
0 & 0 & 1\\
\end{pmatrix},
\label{eq:tilted}
\end{equation}
where $\theta$ and $\phi$ are the pitch and yaw angles, respectively, with positive (negative) values corresponding to right-handed (left-handed) \emph{active} rotations of the triaxial halo. In \appref{rotation} we give detailed derivations. We adopt~\eqref{tilted} for the ellipsoidal radius, \eqref{radius}, throughout the remainder of this paper.

Various studies of the stellar halo and stellar streams have yielded halo models with different values of $\phi$, $\theta$, $p$, and $q$; see~\tabref{titlting} for a summary. These models can be broadly categorized into two groups: (I) those with a small yaw angle ($\phi \sim 20^\circ-30^\circ$) and a small to moderate pitch angle ($\theta \sim 20^\circ-40^\circ$), with flattening parameters $p\approx q$ in the range $0.7-0.9$; and (II) those with a large yaw angle ($\phi \sim 90^\circ-100^\circ$) and a moderate pitch angle ($\theta \sim 40^\circ - 50^\circ$), with $q< p\approx 1$. 
The first group includes the fiducial \emph{Gaia}-Sausage-Enceladus (GSE) model of~\cite{2022AJ....164..249H}, model 1 of~\cite{Nibauer:2025jvo}, the halo model of~\cite{Li:2025oxg}, and the fiducial model of~\cite{2025arXiv251000095D}. The second group includes model 2 of~\cite{Nibauer:2025jvo} and the triaxial model of~\cite{Putney:2025mch}. Based on these findings, we select one representative benchmark model (BM) from each group. For BM I (small yaws), we adopt the fiducial GSE model of~\cite{2022AJ....164..249H}, while for BM II (large yaws), we adopt model 2 of~\cite{Nibauer:2025jvo}. The specific parameters are listed in~\tabref{benchmarks}, and the corresponding 3D views are shown in~\figref{benchmark}. We note that the difference in the yaw angles between the BM I (red) and the BM II (cyan) is $73^{\circ}$. That angle difference can be better seen in~\figref{benchmark} by comparing $b_{\textrm{I}}$ and $b_{\textrm{II}}$ vectors, as those are fixed when we impose an intrinsic pitch.
Since tilted halo studies typically use data from galactocentric distance $r \geq 4-10\,\text{kpc}$, the halo at smaller radii ($r \lesssim 3\,\text{kpc}$) could, in principle, have a different tilt and flattening.\footnote{Stellar tracers respond to the total gravitational potential. Therefore, such deviations must be compensated at the outskirts regions to yield the BMs.} 
To explore this possibility, we consider a variant of BM I, in which the yaw angle is flipped from $24^\circ$ to $-24^\circ$ while all other halo parameters remain fixed. This flipped variant is to illustrate a more diverse tilted halo. We leave a comprehensive study of inferring the yaw, pitch, roll, $p$, and $q$ of the 3D dark matter halo, from the observed GCE morphology to future work. 

\begin{table}[t]
\centering
\topcaption{Parameters for the tilted halo benchmarks used in this work.}
\label{tab:benchmarks}
\begin{tabular}{lccccl}
\hline
BM & Yaw ($\phi$) & Pitch ($\theta$) & $p\equiv b/a$ & $q\equiv c/a$ & Ref. \\
\hline
I  & $24^\circ$ & $25^\circ$ & 0.81 & 0.73 & \cite{2022AJ....164..249H} \\
II & $97^\circ$ & $56^\circ$ & 0.95 & 0.65 & \cite{Nibauer:2025jvo} \\
Flipped I & $-24^\circ$ & $25^\circ$ & 0.81 & 0.73 & --                      \\
\hline
\end{tabular}
\end{table}

To fully specify the density profile in~\eqref{density}, we must also fix the scale radius and density. We set $r_s = 20\,\text{kpc}$ and choose the scale density such that the local dark matter density satisfies~\cite{Catena:2009mf,Salucci:2010qr},
\beq
\rho (r_e (\bm x_\odot))  = 0.4~\text{GeV}/\text{cm}^3.
\label{eq:local}
\eeq
Note that estimates of the local dark matter density depend on the assumed halo model, and a tilted halo can, in principle, yield different values. Reference ~\cite{Putney:2025mch} investigated a tilted halo similar to BM II and found an inferred local dark matter density of $0.32\pm 0.03~\text{GeV}/\text{cm}^3$, which is not significantly different from~\eqref{local}. Since BM II represents the more extreme tilt of the two benchmarks, we expect the inferred local dark matter density for (flipped) BM I to be even closer to~\eqref{local}. We therefore adopt~\eqref{local} for all the benchmarks. Also note that changes in the local dark matter density will act as an overall scaling of the GCE template. While it will change the inferred dark matter annihilation cross section, it does not affect the inferred halo morphology. 

\begin{figure*}[t]
   \centering
   \includegraphics[width=0.96\textwidth]{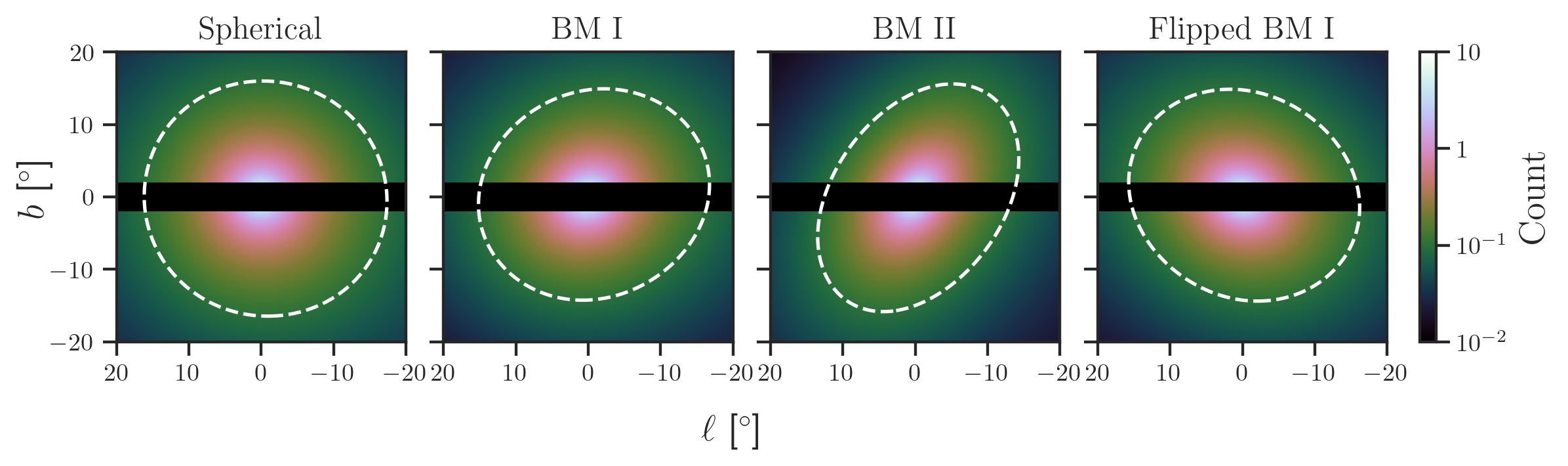}
   \caption{GCE count maps for a spherical dark matter halo (left), tilted halo BM I (middle left), tilted halo BM II (middle right), and the flipped BM I (right). For all models, we assume a cuspiness of $\gamma = 1.2$, take the energy range of $1.02-1.32$ GeV, and mask the disk region with $|b| < 2^\circ$. The white dashed lines show contours of equal count to guide the eye.}
   \label{fig:example}
\end{figure*}

Figure \ref{fig:example} shows the GCE count maps for a spherical halo, BM I and II, and the flipped BM I halo models, all with cuspiness $\gamma = 1.2$ at the energy of $1.02-1.32$ GeV.  Positive yaw and pitch angles (BM I and II) correspond to right-handed rotations of the major axis, causing the resulting GCE to elongate along the $(-\ell, +b)$ to $(+\ell, -b)$ direction. The morphology of (flipped) BM I does not differ significantly from that of a spherical halo, whereas BM II more closely resembles a prolate GCE due to its small $q$ and large yaw and pitch angles.

\section{The Fitting Procedure}
\label{sec:fitting}

We utilize \textit{Fermi}-LAT Pass 8 data \cite{Fermi-LAT:2013jgq}, collected between August 4, 2008 (week 9) and April 29, 2025 (week 882). \textit{Fermi} \texttt{ScienceTools P8v2p4p0} are used for selection cuts and to generate the exposure maps. We restrict our analysis to the CLEAN event class, using only FRONT-converting events for better angular resolution. 
We also apply the following filtering procedure to ensure good data quality: \(z_{\text{max}} = 100^\circ\), \(\texttt{DATA\_QUAL} = 1\), \(\texttt{LAT\_CONFIG} = 1\), and \(|\texttt{ROCK\_ANGLE}| < 52^\circ\). The region of interest (ROI) is set as the inner 40 degrees of the GC in the heliocentric Galactic coordinates ($-20\degree \leqslant \ell \leqslant 20\degree$, $-20\degree \leqslant b \leqslant 20\degree$ ) and we bin the ROI in Cartesian pixels of size $0.1\degree \times 0.1\degree$. Additionally, the data are binned in 14 logarithmically spaced energy bins between $0.275$ GeV $\leqslant E_{\gamma} \leqslant 51.9$ GeV (see~\tabref{mask}). 
In addition, we mask all the point sources from the 4FGL-DR4 catalog~\cite{Ballet:2023qzs} within the ROI with an energy-dependent mask radius based on having either a low or high test statistic (TS $< 49$ or $\geq 49$), as shown in~\tabref{mask}, and the Galactic disk within $|b| < 2^{\circ}$. We refer to that mask as ``4FGLDR4 + L20.''

Next, for each energy bin $j$, we construct the total diffuse flux map in the $40\degree \times 40\degree$ ROI by modeling the flux in each pixel $p$ as a weighted sum of the individual diffuse components,  
\beq
\Phi_{j,p}^{\rm TOT} =\sum_{T} c_{j}^{T} \Phi_{j,p}^{T},
\eeq 
where the template index $T$ runs over the $\pi^0$, ICS, bremsstrahlung, \emph{Fermi} bubbles, and isotropic templates. When fitting models that include the GCE, $T$ also includes the GCE template. The coefficients $c_{j}^{T}$ are normalization factors for each template. Because the $\pi^0$ and bremsstrahlung templates have similar (but not identical) spatial distributions, we impose a shared normalization, $c_j^{\rm Gas} \equiv c_j^{\pi^0}=c_j^{\rm Bremss}$. The corresponding count map is then obtained by multiplying the total flux map with the exposure map, ${\cal C}_{j,p} = {\cal E}_{j,p} \Phi_{j,p}^{\rm TOT}$.

Given the expected count map $\mathcal C_{j,p}$, the observed \emph{Fermi}-LAT count map $\mathcal D_{j,p}$, and the mask $\mathcal M_{j,p}$, the total log-likelihood $\ln (\mathcal L)$ is constructed as $-2 \ln   (\mathcal L) = -2 \sum_j \ln (\mathcal L_j)$ with,
\begin{widetext}
\begin{equation}
\label{eq:log-likelihood_function}
-2\ln ({\cal L}_j) = \Bigg\{ 2\sum_{p}\Bigg[{\cal M}_{j,p}{\cal C}_{j,p} + \ln\left[({\cal M}_{j,p}{\cal D}_{j,p}\right)!]  - {\cal M}_{j,p}{\cal D}_{j,p} \ln\left({\cal M}_{j,p}{\cal C}_{j,p}\right) \Bigg]\Bigg\} + \chi^{2}_{{\rm Bubbles},j} + \chi^{2}_{{\rm Iso},j}.
\end{equation}
\end{widetext}
where $\chi^{2}_{{\rm Bubbles},j}$ and $\chi^{2}_{{\rm Iso},j}$ are penalty terms, constraining the normalizations of the \emph{Fermi} Bubbles and isotropic components to remain consistent with their measured spectra at high latitudes, as described in Refs.~\cite{Fermi-LAT:2014sfa, Fermi-LAT:2014ryh}. 

We sample the posterior distribution of the normalization coefficient $c_{j}^{T}$ using \texttt{DYNESTY}~\cite{Speagle:2019ivv}, a nested-sampling method, and then refine the inference with a NUTS/HMC sampler of \texttt{NumPyro}~\cite{Phan:2019elc}, within the analysis pipeline \texttt{gcepy}~\cite{McDermott:2022zmq, Cholis:2021rpp}. We analyze the resulting chains using \texttt{ChainConsumer}~\cite{2016JOSS....1...45H}, to estimate the posterior distributions of $c_{j}^{T}$ (central values and $1\sigma$ uncertainties). We also use \texttt{scipy.optimize.differential\_evolution}~\cite{Virtanen:2019joe}, a stochastic optimizer, to verify that $-2\ln \mathcal L_j$ from the \texttt{HMC} sampler converges to the global minimum. We find that the differences between the \texttt{HMC} sampler and \texttt{differential\_evolution} are $|-2\Delta\ln \mathcal L_j| \lesssim  \mathcal{O}(1)$.

\section{Results}
\label{sec:results}

We first test the hypothesis that the GCE is the annihilation signal from a triaxial dark matter halo that is tilted and rotated with respect to the Sun-Galactic Center axis. As described in \secref{setup}, we consider tilted halo BM I with cuspiness in the range of $0.9 \leq \gamma \leq 1.5$. A value of $\gamma = 1.0$ represents the conventional choice for an NFW profile~\cite{Navarro:1995iw}, while a choice of $\gamma = 1.2-1.3$ is the typical best-fit result for the GCE assuming a spherical dark matter annihilation signal (see Refs.~\cite{Daylan:2014rsa, Calore:2014xka, DiMauro:2021raz, Cholis:2021rpp, Zhong:2024vyi}).

\begin{figure}
    \centering
    \includegraphics[width=1.0\linewidth]{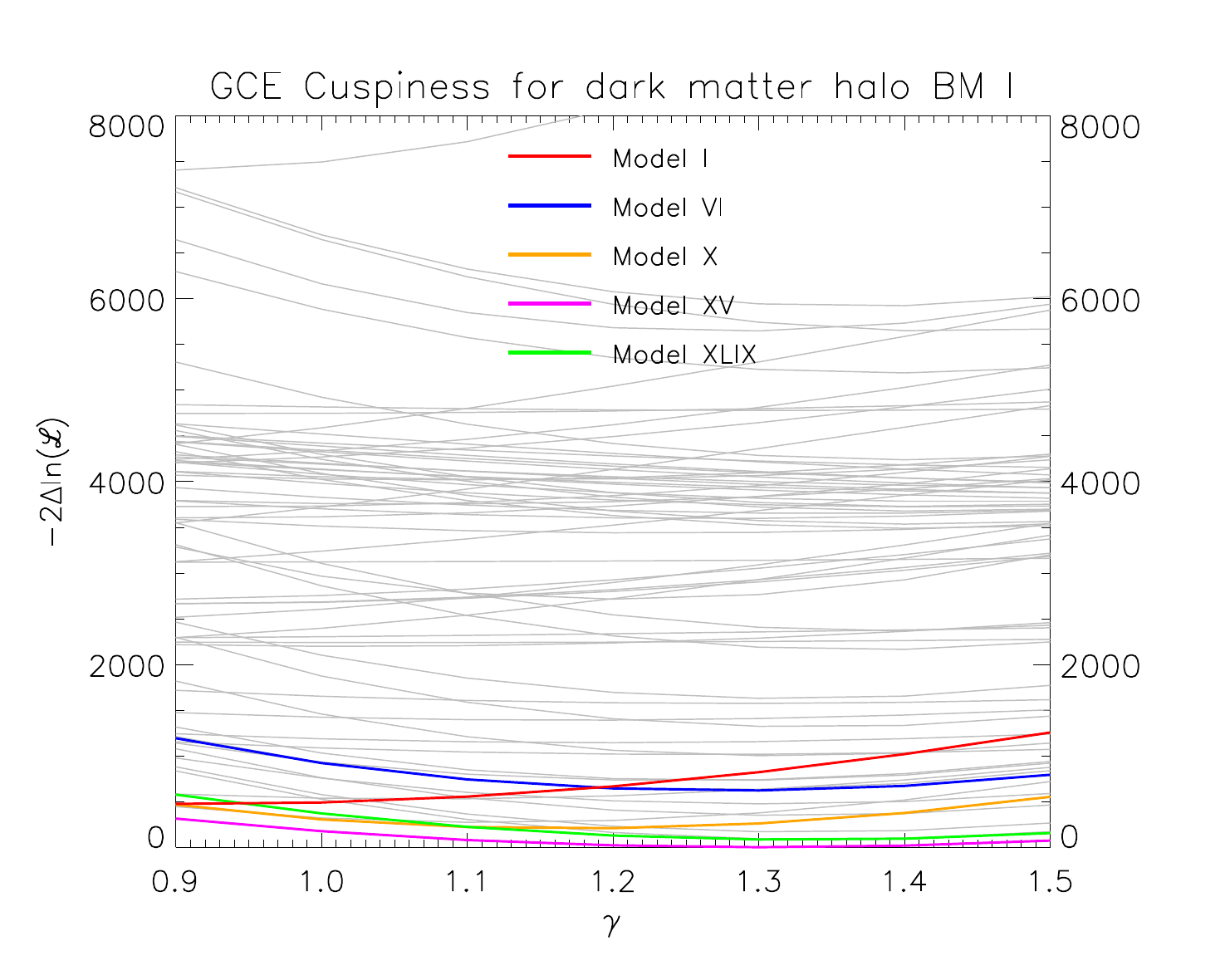}
    \vspace{-0.8cm}
    \caption{The cuspiness $0.9 \leq \gamma \leq 1.5$, of the GCE morphology for a generalized NFW triaxial dark matter annihilation profile tilted to our line of sight, for the assumptions of Ref.~\cite{2022AJ....164..249H}, i.e., BM I of Fig.~\ref{fig:example}. The best-fit GDE models give a preference for $\gamma\simeq 1.2-1.3$. 
    Our $y$-axis gives the difference in the fit between choices for the GCE morphology, defined as $-2 \ln \mathcal{L}$ for the alternative model minus $-2 \ln \mathcal{L}$ of the best fit GCE from the entire sample. Colored lines showcase the results for some of the GDE models that give the best fit, while the gray lines show the GCE under every single GDE model.}
    \label{fig:GCE_cuspiness}
\end{figure}

Figure~\ref{fig:GCE_cuspiness} shows the resulting difference, $2 \Delta \ln (\mathcal L)$, between the log-likelihoods of 80 GDE models of  GDE~\cite{Cholis:2021rpp} and that of the best-fit GDE model, as a function of  $\gamma$, coloring the results for some of the GDE models that give the best-fit results (lower lines along the $y$-axis).  
We find that the GDE models with the best fit give a preference for $\gamma = 1.2$ or $1.3$ as has been previously claimed in the literature. Among the models that give good fits, an exception is GDE model I (red line), which gives its best fit for $\gamma = 0.9-1.0$. In all of those cases, the best-fit cuspiness values are consistent with expectations for a dark matter density profile to have as slope in its inner $\simeq 3$ kpc that our gamma-ray fits probe. Among the 80 GDE models tested, only a small number that in fact give the worst-fit results prefer a value of $\gamma = 1.4$ or $1.5$; i.e., values that would be extreme assumptions for the inner slope of a dark matter density profile.

We also tested our BM II from Ref.~\cite{Nibauer:2025jvo}. 
We find that even for this large yaw and pitch profile, there is a preference for the cuspiness values in the range of $\gamma \simeq 1.2-1.4$. BM I is preferred to BM II by $2 \Delta \ln(\mathcal{L}) \simeq (5-8)\times 10^{2}$ when using the same GDE models. 
We present the relevant results for the dark matter annihilation profile's cuspiness for BM II in Appendix~\ref{app:AlternativeDMprofile_Nibauer_Bonaca}.

\begin{figure}
    \centering \includegraphics[width=1.0\linewidth]{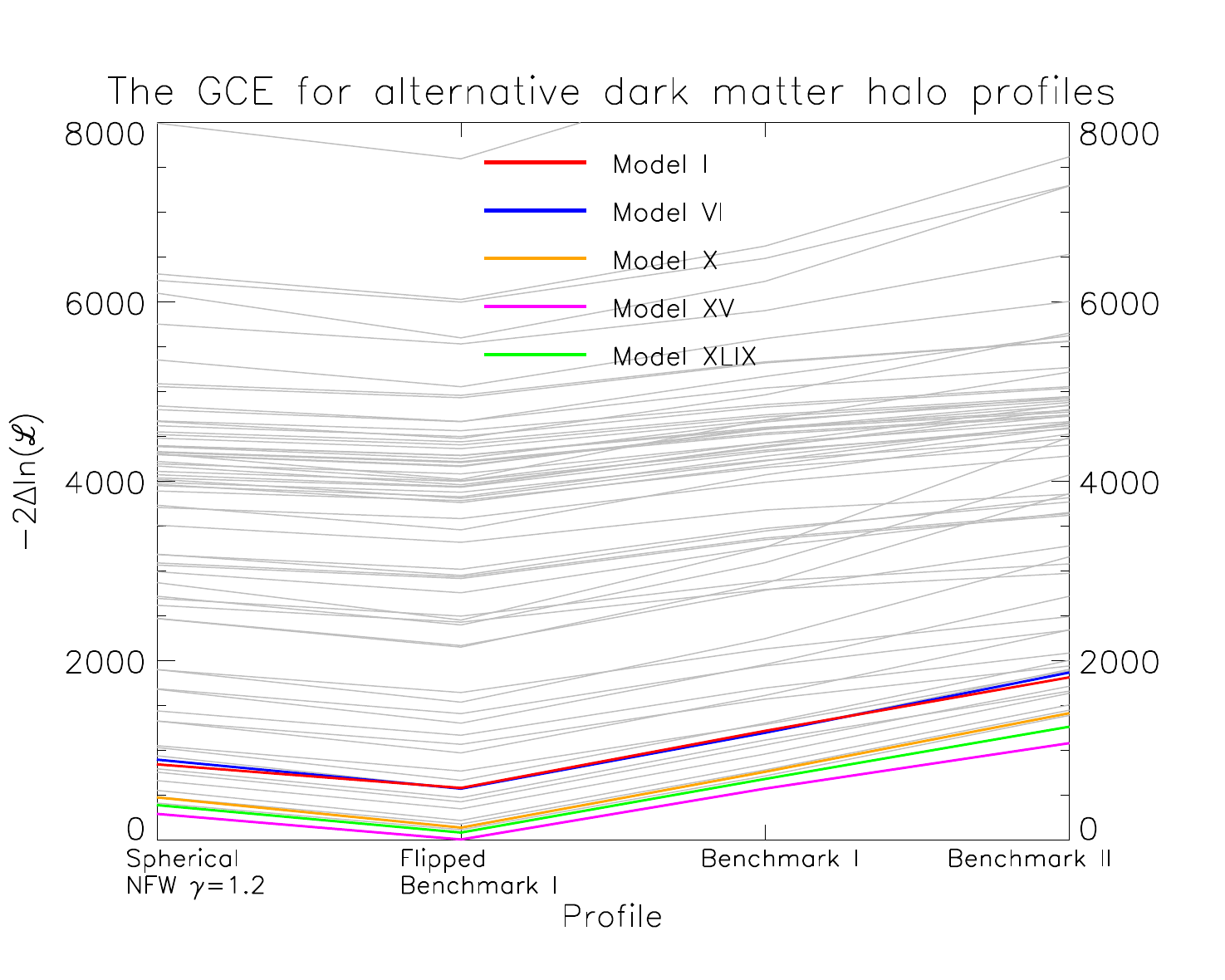}
    \vspace{-0.8cm}
    \caption{The morphology of the GCE for all 80 GDE models. We compare for an inner slope $\gamma=1.2$ of the dark matter halo profile, the spherical GCE template to those of BM I, BM II and the flipped BM I of~\figref{example}. We find that there is a systematic preference for the flipped benchmark I assumptions.}
    \label{fig:GCE_for_alternative_profiles}
\end{figure}

Interestingly, we find that the \textit{Fermi}-LAT observations give a better fit for the flipped BM I (shown in the rightmost panel of Fig.~\ref{fig:example}), which is preferred over the regular BM I by $2 \Delta \ln(\mathcal{L}) \simeq 6\times10^{2}$.
Furthermore, the spherically symmetric dark matter halo option, as shown in the leftmost panel of Fig.~\ref{fig:example}, is preferred over BM I by $2 \Delta \ln(\mathcal{L}) \simeq (1.5-2)\times10^{2}$.  
All these results are shown for all 80 GDE models in Fig.~\ref{fig:GCE_for_alternative_profiles}, for the case of $\gamma=1.2$ (and in Appendix~\ref{app:AlternativeDMprofile_Nibauer_Bonaca} for $\gamma=1.0$).
The statistical preference for the flipped BM I, followed by the preference for a spherical profile, followed by the BM I, with the BM II being the least preferred, is consistently present independent of the GDE model and found also when using for the inner slope/cuspiness a value of $\gamma=1.0$ and of $\gamma=1.3$. 
We remind the reader that the halo profiles of Refs.~\cite{2022AJ....164..249H, Nibauer:2025jvo} are inferred from the stellar tracers at distances of $r \geq 6\,\text{kpc}$ or $10\,\text{kpc}$. The dark matter profile based on kinematical observations in the inner kiloparsecs of our Galaxy is still quite uncertain.  

\begin{figure}
    \centering
    \includegraphics[width=1.0\linewidth]{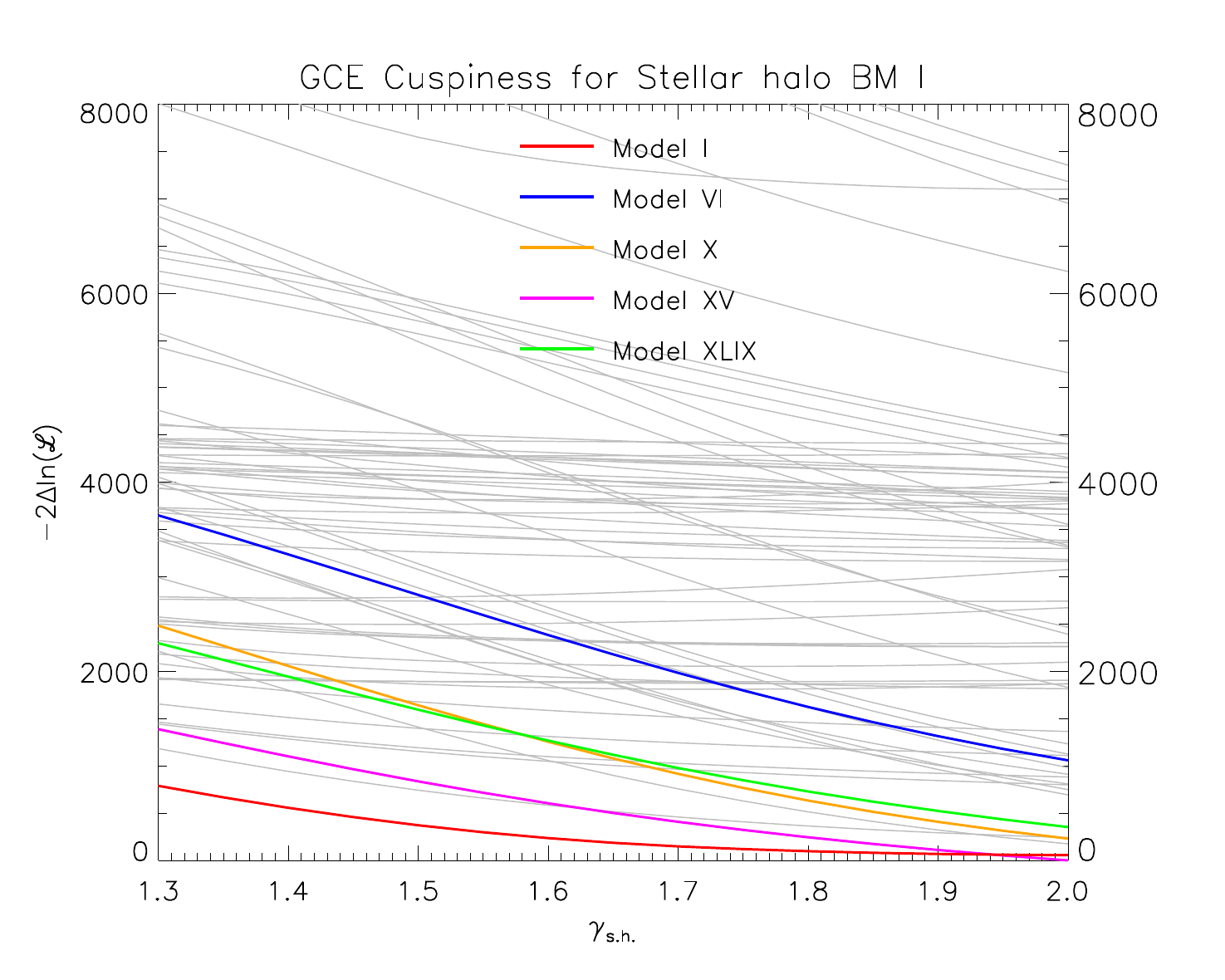}
    \vspace{-0.8cm}
    \caption{We assume that the GCE follows the stellar halo profile with cuspiness of $1.3 \leq \gamma_{\textrm{s.h.}} \leq 2.0$. Like in Fig.~\ref{fig:GCE_cuspiness}, we assume that the profile is triaxial and tilted to our line of sight, for the assumptions of Ref.~\cite{2022AJ....164..249H}. The best-fit results go beyond what is considered the expected range of cuspiness for a stellar halo. Our $y$-axis gives the difference in the fit $-2 \Delta \ln(\mathcal{L})$ between choices for the GCE morphology. Colored lines show the results for some of the best-fit GDE models, while the gray lines show the GCE under every single GDE model.}
    \label{fig:GCE_cuspiness_stellar_halo}
\end{figure}

We also test the case that the GCE is correlated to the stellar halo,\footnote{
Note that the stellar halo-motivated GCE template used here are different from the stellar bulge/bar template motivated by primordial or dynamical formation channels for millisecond pulsars~\cite{Macias:2019omb}, which suggest a gamma-ray emissivity scale with the boxy-bulge density by $j_\gamma \propto \rho_\text{bar}^s$ with $s\approx 1.38$. } a signature that would suggest its origin is more likely the gamma-ray emission of a population of millisecond pulsars. We take the stellar halo to also have an inner slope/cuspiness $\gamma_{\textrm{s.h.}}$, with $1.3 \leq \gamma_{\textrm{s.h.}} \leq 2.0$.\footnote{While the dark matter signal is one of annihilation and thus the signal scales as $\rho_\text{DM}^2$, i.e., the exponent of its morphology is equal to $2 \gamma$, for the stellar halo that exponent is still the $\gamma_{\textrm{s.h.}}$.} Again, we impose $\phi = 24^{\circ}$ and $\theta = 25^{\circ}$ rotations and assume that the stellar halo is also triaxial with $p = 0.81$ and $q = 0.73$ that are the assumptions for BM I. We show our results in Fig.~\ref{fig:GCE_cuspiness_stellar_halo}.

For the assumption that the GCE has a profile following the stellar halo, we find that the best fit for $\gamma_{\textrm{s.h.}}$ is 2.0 or larger, or that its value is very difficult to constrain (flat gray lines). Values of $\gamma_{\textrm{s.h.}} \geq 2.0$ are mostly disfavored by studies of the stellar halo's morphology (see e.g. Refs~\cite{2022AJ....164..249H, Li:2025oxg} and their respective Table I). 
We note that the best-fit stellar halo assumptions for the GCE (models I and XV) give a similar quality of fit to the best-fit models under the assumption of the GCE following a dark matter annihilation profile (several lines in Fig.~\ref{fig:GCE_cuspiness}, including models I, X, XV, and XLIX). However, we point out that such GDE models achieve this only for $\gamma_{\textrm{s.h.}} \geq 1.8$ and only for finely tuned GDE model assumptions. 
Furthermore, we note that for values of $\gamma_{\textrm{s.h.}} \simeq 2$, the stellar halo acquires a morphology that is very similar to that expected from a dark matter annihilation profile. At these values for $\gamma_{\textrm{s.h.}}$, we are effectively defining as stellar halo a profile that is indistinguishable from that of dark matter annihilation, while also being in tension with the stellar halo's density profile as inferred from stellar kinematics. We find the same results when we test the flipped BM I for the stellar halo. 

In our opinion, the results of Figs.~\ref{fig:GCE_cuspiness} and~\ref{fig:GCE_cuspiness_stellar_halo} already show a preference for the GCE profile that is consistent with the expected morphology from dark matter annihilation and not that of a stellar halo. 

\begin{figure}
    \centering
    \includegraphics[width=1.0\linewidth]{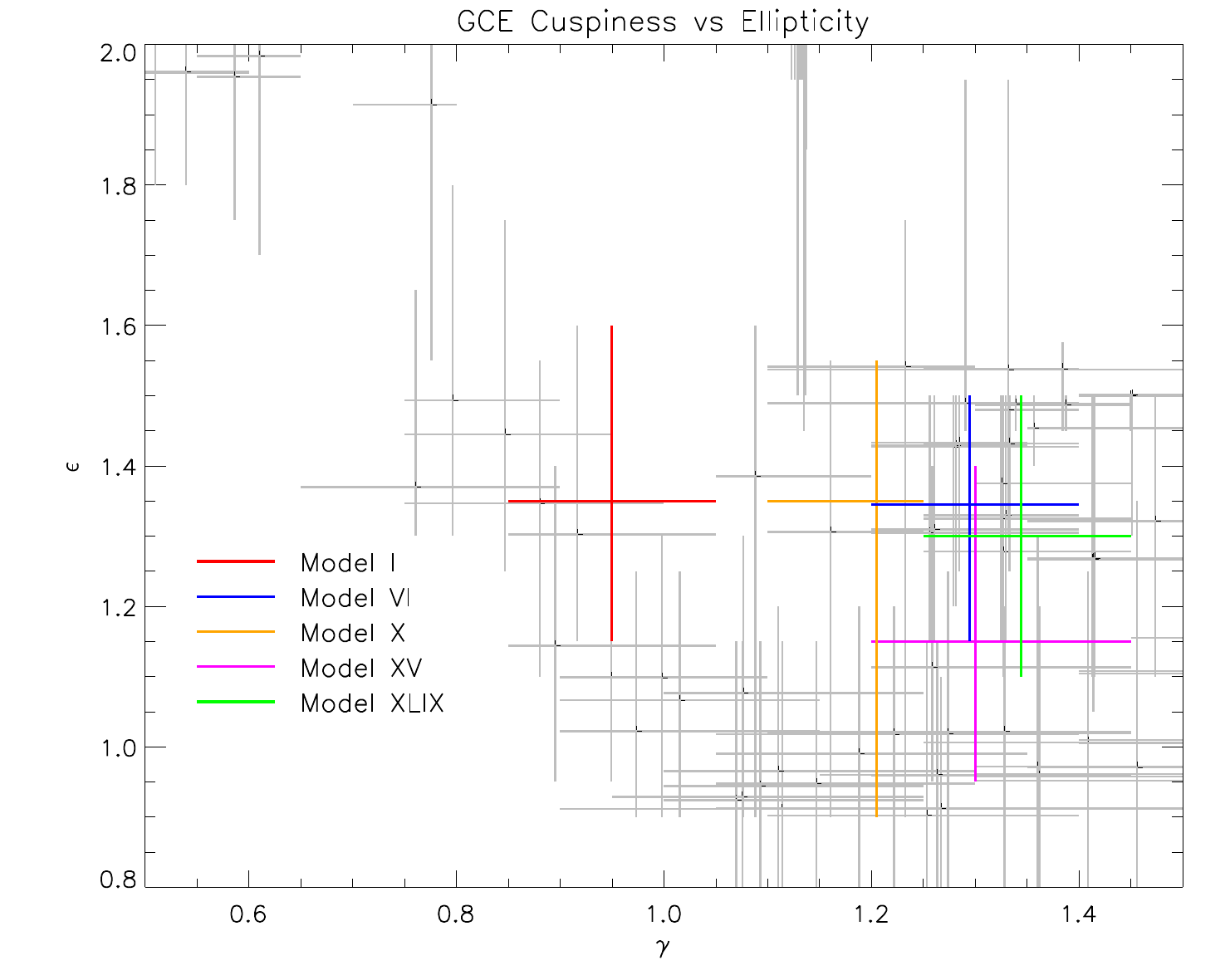}
    \caption{The best fit ranges of the GCE's cuspiness $\gamma$ and ellipticity $\epsilon$ for all 80 diffuse emission models. We assume \emph{no yaw or pitch} on these profiles. The error bars around each point give the ranges of $\gamma$ and $\epsilon$ for a given GDE model that are within $-2\Delta \ln(\mathcal{L}) \leq 50$ from its best-fit choice. In color, we showcase the results for some of the diffuse emission models that give the best overall fit. To avoid overlap of the crosses of different diffuse emission models, we have imposed a small random displacement of the best-fit points.}
    \label{fig:cuspiness_vs_ellipticity}
\end{figure}

In Fig.~\ref{fig:cuspiness_vs_ellipticity}, we show for each of the 80 GDE models the GCE's best-fit ranges for its cuspiness $\gamma$ and ellipticity $\epsilon$, for a dark matter halo without any tilt. The ellipticity $\epsilon$
is used to modulate the opening angle from the GC $\psi$ of the dark matter template, in such a manner that $cos(\psi) = cos(b)cos(\ell/\epsilon)$. The error bars represent the ranges of $\gamma$ and $\epsilon$ that are within a $-2\Delta \ln(\mathcal{L}) \leq 50$ from the best-fit ($\gamma$, $\epsilon$) combination of each GDE model. 
Again, we showcase in color the results for five among the GDE models that give the overall best-fit results. Except for a small number of GDE models, there is a systematic preference for $0.9 \leq \gamma \leq 1.5$ and $ 0.8 \leq \epsilon \leq 1.5$. The GDE models that give high values of $\epsilon$ also give $\gamma \simeq 0.6$ (i.e., they are at the top left corner of Fig.~\ref{fig:cuspiness_vs_ellipticity}), and are among the models that have the worst overall fit (as are models XLIV, XLV, XLVI of~\cite{Cholis:2021rpp}).

\begin{figure}
    \centering
    \includegraphics[width=1.0\linewidth]{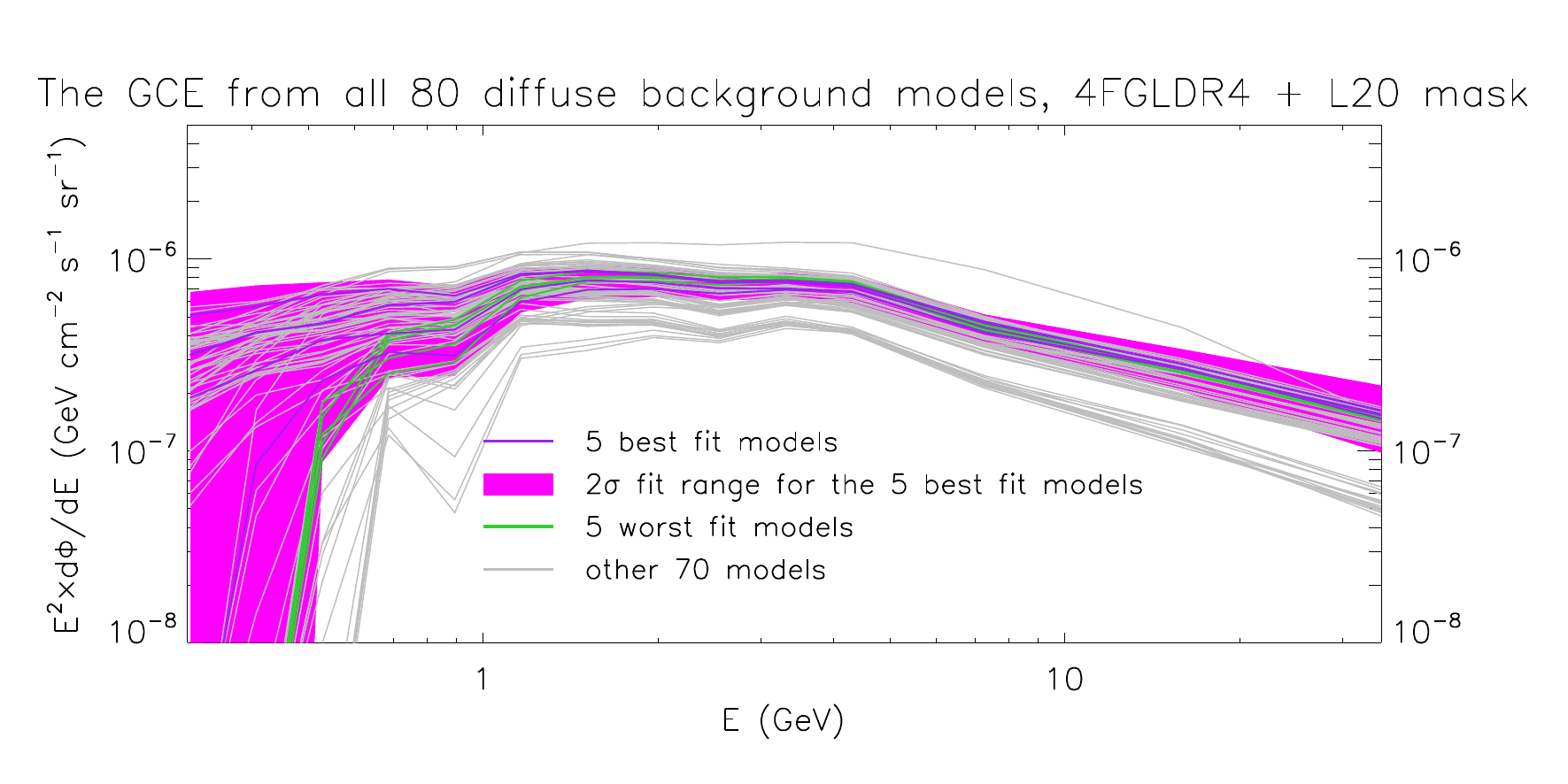}
    \caption{The GCE spectrum in our $40^{\circ}\times 40^{\circ}$ masked region of interest, assuming a triaxial and tilted dark matter profile with the assumptions of Ref.~\cite{2022AJ....164..249H} (BM I), with $\gamma=1.2$. The GCE flux is allowed to be negative in the fit, which occurs for a few models at energies below 0.7 GeV. The magenta lines show the best-fit normalization of the GCE for the five GDE models that give the best overall fit to the \textit{Fermi}-LAT data. In the magenta band, we plot their combined $2\sigma$ uncertainty range. The green lines give the GCE spectrum from the five diffuse emission models that have the worst fit to the \textit{Fermi}-LAT data. The gray lines give the GCE spectrum from the remaining 70 out of the 80 GDE models used. The GCE spectrum from the 80 diffuse emission models is effectively identical for a triaxial and tilted dark matter profile with either $\gamma=1.1$, $\gamma=1.3$, or $\gamma=1.4$.}
    \label{fig:GCE_spectrum}
\end{figure}

In Fig.~\ref{fig:GCE_spectrum}, we plot the resulting GCE spectra from our fit to the $40^{\circ}\times 40^{\circ}$ ROI.  
We plot all 80 GCE spectra that come from using each of the 80 GDE models of Ref.~\cite{Cholis:2021rpp}. 
We indicate in the magenta and in the green lines, the GCE spectra that are evaluated using the GDE models that give the five best-fit results (in magenta) and the five worst-fit results (in green) to the \textit{Fermi}-LAT data. 
The GDE models with the best-fit results to the data are models (starting from the very best result) XV, XLVIII, XLIX, XLVII, and X. While the five models that give the worst fit are LXII, LXIII, LXX, LXXI, and LXIV. 
In Fig.~\ref{fig:GCE_spectrum}, for the GCE template, we use the triaxial and tilted dark matter annihilation case of BM I with $\gamma=1.2$. 
Had we used the $\gamma=1.1$, $\gamma=1.3$ or $\gamma=1.4$ equivalent triaxial and tilted template, the spectral results would have been effectively (within a few $\%$) identical; while even a non-triaxial or rotated, i.e., spherical NFW-like dark matter annihilation template with $\gamma \simeq 1.2$, or the flipped BM I, give effectively the same GCE spectra.
Even the BM II gives almost the same overall GCE spectrum, from the $40^{\circ} \times 40^{\circ}$ ROI, as presented in Fig.~\ref{fig:GCE_spectrum}.  
We present the relevant results for the dark matter annihilation profile in Appendix~\ref{app:AlternativeDMprofile_Nibauer_Bonaca}.
Irrespective of the exact triaxial or spherical shape of the dark matter profile (from the published literature) and its exact set of rotations, the GCE spectrum is essentially the same for $\gamma \simeq 1.1-1.4$ as presented in Fig. 3 of Ref.~\cite{Zhong:2024vyi} and Fig. 12 of Ref.~\cite{Cholis:2021rpp}, showcasing the robustness of the GCE spectral properties to the continuous observations from the \textit{Fermi}-LAT and the increasing number of newly identified galactic gamma-ray point sources, that are masked using the gradually updated catalogs provided by the Fermi collaboration.\footnote{In this work we used the 4FGL-DR4 gamma-ray point source catalog, while in Ref.~\cite{Cholis:2021rpp} the then available 4FGL-DR2 catalog was used. In Ref.~\cite{Zhong:2024vyi}, the 4FGL-DR3 was used as a basis with multiple alternative masking choices being tested.}

\section{Conclusion and Discussion}
\label{sec:discussion}

In this work, we have investigated how the triaxiality and tilt of the MW dark matter halo affect the properties of the Galactic Center Excess. Our analysis extends previous GCE studies by considering (1) untilted triaxial halos with a large set of varying ellipticity and cuspiness combinations, encompassing the entire range of these morphological parameters for the GCE and (2) a set of specific tilted dark matter halo configurations motivated by recent stellar halo and stellar stream studies.

A central result of our analysis is the robustness of the GCE spectrum and cuspiness. Across all halo configurations considered, including the spherical and tilted scenarios (BM I, II, and the flipped BM I), we consistently find that the best-fit cuspiness is $\gamma \approx 1.2-1.3$, in agreement with previous studies~\cite{Calore:2014xka, Daylan:2014rsa, TheFermi-LAT:2015kwa, Cholis:2021rpp, Zhong:2024vyi}. The inferred spectrum is also robust, varying by only $\sim 10\%$ in its overall amplitude (but only within a few $\%$ in its shape) across different halo configurations and for $\gamma \approx 1.1-1.3$, for any given GDE model.

While the spectrum and cuspiness are robust, the analysis does exhibit some discriminating power among 3D configurations of the dark matter halo. We find a systematic preference for the flipped BM I, followed by the spherical profile, then BM I, with BM II, whose shape is close to prolate, being the least favored. These preferences are consistent across a collection of various best-fit galactic diffuse emission models and cuspiness values, suggesting that the GCE morphology could be used to infer the inner halo structure.

The GCE morphology also disfavors a stellar halo origin with a 3D morphology as suggested by Ref.~\cite{2022AJ....164..249H} (fiducial GSE model). The GCE's morphology being correlated with known stellar populations is a prediction of the MSP interpretation for its origin. When we fit the GCE with the stellar halo profile, the best-fit cuspiness is $\gamma_{s.h.} \geq 2.0$, which is in tension with the cuspiness of the stellar halo inferred from stellar kinematics. With such steep values, the stellar halo profile becomes indistinguishable from dark matter annihilation (since the annihilation signal scales as $\rho^2$). 

Several caveats apply. The tilted halo models we studied are limited. Most of them use stellar kinematic measurements at $r \geq 4-10\, \text{kpc}$, and the actual configuration in the GCE region ($r\lesssim 3\,\text{kpc}$) could be more complicated. For example, the ellipsoidal shells may exhibit twisted isodensity contours (see simulation results in~\cite{Hussein:2025xwm}), instead of aligned, which we assume. Nonetheless, a multiwavelength/messenger approach combining gamma-ray and stellar kinematic data offers a promising path forward to jointly constrain the halo structure and the properties of the GCE.

\begin{acknowledgments}
We thank Ana Bonaca, Adam Dillamore, Songting Li, Moorits Mihkel Muru, Jacob Nibauer, Eric Putney, David Shih, and Joseph Silk for useful discussions. L.H. and Y.Z. are supported by the GRF Grants No. 11302824, No. 11310925, and the France/Hong Kong Joint Research Scheme F-CityU106/25 from the Research Grants Council, University Grants Committee, and the Grants No. 9610645 and No. 7020130 from the City University of Hong Kong. I.C. acknowledges that this material is based upon work supported by the U.S. Department of Energy, Office of Science, Office of High Energy Physics, under Award No. DE-SC0022352. Y.Z. acknowledges the Aspen Center for Physics for their hospitality during the completion of this study, which is supported by NSF Grant No. PHY-2210452 and a grant from the Simons Foundation (1161654, Troyer).

\end{acknowledgments}

\section*{Data availablity}
The DM halo maps used in the paper that support the findings of this article are openly available~\cite{github:gce_models}.

\appendix

\section{PROCEDURES OF 3D ROTATIONS}
\label{sec:rotation}

The triaxial halo model and the corresponding density profiles are described in~\secref{setup}. Here we provide more details on the 3D rotation that maps an untilted halo,
\beq
 \bm x^\intercal \bm A \bm x = r_e^2, 
\eeq
where $\bm x = (x, y, z)^\intercal$, to a tilted halo,
\beq
 \bm x'^\intercal \bm A \bm x' = r_e^2~\quad\text{with}~\quad\bm x' = \bm R\, \bm x,
 \label{eq:tilted_halo}
\eeq
where $\bm x' = (x', y', z')^\intercal$ and $\bm R$ is the 3D rotation matrix. The operation is implemented as a sequence of noncommuting rotations, and often different rotation sequences can lead to the same final orientation. We also assume that the halo has reflection symmetries about the mid-planes perpendicular to its principal axes. In other words, we do not distinguish between the ``head'' and ``tail'' of the ellipsoid.

Any tilted configuration can be obtained from an untilted configuration by applying three \emph{active} rotations about the body's principal axes, that is, an \emph{intrinsic} rotation sequence. Under the Tait-Bryan convention, we write the rotation sequence as, 
\beq
z[\phi]\to y'[\theta]\to x''[\zeta].
\label{eq:intrinsic}
\eeq
This means that (1) we first apply a yaw about the minor axis with angle $\phi$, with positive (negative) angles defined by the right-(left-)hand rule, (2) then apply a pitch about the yawed intermediate axis with angle $\theta$, and (3) finally apply a roll about the yawed-then-pitched major axis with angle $\zeta$. The primes and double-primes on $y$ and $x$ indicate that these rotations are defined with respect to the body-fixed axes rather than the external axes.

The same final configuration of~\eqref{intrinsic}, can also be obtained through a sequence of \emph{active} rotations with respect to the external galactocentric coordinate, namely an \emph{extrinsic} rotation,
\beq
X[\zeta]\to Y[\theta]\to Z[\phi],
\label{eq:extrinsic}
\eeq
that is, first applying a roll about the $X$ axis by $\zeta$, then a pitch about the $Y$ axis by $\theta$, and finally a yaw about the $Z$ axis by $\phi$. To show the equivalence between~\eqref{intrinsic} and~\eqref{extrinsic}, we use the pullback relations $\bm R_{y'} = \bm R_z \bm R_Y \bm R_z^{-1}$, $\bm R_{x'} = \bm R_z \bm R_X \bm R_z^{-1}$,
$\bm R_{x''} = \bm R_{y'} \bm R_{x'} \bm R_{y'}^{-1}$, 
together with the identity $\bm  R_z =\bm  R_Z$. It then follows that~\eqref{intrinsic} corresponds to the rotation matrices, 
\begin{align}
\bm  R_{x''} \bm  R_{y'} \bm  R_z ={}& (\bm  R_y' \bm  R_x' \bm  R_y'^{-1}) \bm  R_{y'} \bm  R_z \nonumber\\
={}& \bm  R_Z \bm  R_Y \bm  R_Z^{-1} \bm  R_Z \bm  R_X \bm  R_Z^{-1} \bm  R_Z = \bm  R_Z \bm  R_Y \bm  R_X,
\end{align}
where the right-hand side is precisely the rotation matrices corresponding to~\eqref{extrinsic}.

We can equivalently recast the operation in terms of \emph{passive} rotations, in which the halo remains fixed while the galactocentric coordinate system $XYZ$ is rotated, resulting in an apparently tilted halo. Under a \emph{passive} rotation  $\bm{R}$, the coordinates of a point in the new (rotated) frame, $\bm{x}'$, are related to its coordinates in the original frame, $\bm{x}$, by $\bm{x}' = \bm{R}\, \bm{x}$. This coordinate transformation is directly related to the analytical expression for a tilted halo in~\eqref{tilted_halo}.

Because a passive rotation of the coordinate system is the \emph{inverse} of an active rotation of the object, the following operations on the $(X, Y, Z)$ axes,  
\beq
\text{(Passive~Rotation) } Z\{-\phi\}\to Y\{-\theta\}\to X\{-\zeta\},
\label{eq:extrinsic2}
\eeq 
are equivalent to the active rotation of~\eqref{extrinsic}.
Since~\eqref{intrinsic} is equivalent to~\eqref{extrinsic}, 
the passive rotation of~\eqref{extrinsic2} is also equivalent to the intrinsic active rotation of~\eqref{intrinsic}. In other words, to implement the Tait-Bryan sequence $z\to y'\to x''$ for the triaxial halo, the rotation matrix $\bm{R}$ for~\eqref{tilted_halo} should be the rotation matrix corresponding to~\eqref{extrinsic2},
\beq
\bm{R} = \bm{R}_{X}(-\zeta)\bm{R}_{Y}(-\theta)\, \bm{R}_{Z}(-\phi),
\label{eq:rot_mater}
\eeq
where $\bm R_Z$, $\bm R_Y$, $\bm R_X$ are respectively given by,
\begin{align}
\bm{R}_{Z}(-\phi) ={}& \begin{pmatrix}
\cos\phi & \sin \phi & 0\\
-\sin \phi & \cos \phi & 0\\
0 & 0 & 1
\end{pmatrix},
\label{eq:rot1}
\\
\bm{R}_{Y}(-\theta) ={}& \begin{pmatrix}
\cos \theta & 0 & -\sin\theta\\
0 & 1 & 0\\
\sin \theta & 0 & \cos \theta
\end{pmatrix}, 
\label{eq:rot2}
\\
\bm{R}_{X}(-\zeta) ={}& \begin{pmatrix}
1 & 0 & 0\\
0 & \cos \zeta & \sin \zeta\\
0 & -\sin \zeta & \cos \zeta
\end{pmatrix}.
\label{eq:rot3}
\end{align}
The signs of $\sin \phi$ terms in~\eqref{rot1} are chosen so that $\bm{R}_{Z}(\alpha>0)$ or $\bm{R}_{Z}(\alpha<0)$ corresponds to a right-handed or left-handed rotation, respectively. We adopt the same sign convention for $\theta$ and $\zeta$ and choose the signs of sine terms in~\eqref{rot2} and~\eqref{rot3} accordingly.
In the special case where the roll is neglected ($\zeta = 0$), so that only yaw-then-pitch is applied,~\eqref{rot_mater} reduces to~\eqref{tilted}.

\section{TRIAXIAL AND TILTED HALO IN LITERATURE}
\label{sec:survey}

 \begin{table*}[t]
     \centering
        \topcaption{Summary of models of the tilted triaxial halo in Group I (small yaws, upper panel) or II (large yaws, lower panel) halo in literature. For each model, we provide the minimum and maximum of the Galactocentric distance of the model data ($r_{\min}$ and $r_{\max}$), the flattening parameters ($p$ and $q$), the 3D rotation operation, and the inner halo cuspiness, $\gamma_\text{inner}$. }
     \renewcommand{\arraystretch}{1.5} 
     \begin{tabular}{c c | c | c c c c}
     \hline
          Ref. & Model  & $r_{\min}-r_{\max}$ [kpc] & $p\equiv b/a$ & $q \equiv c/a$ & Rot. Axis~[angles] & $\gamma_\text{inner}$ \\
     \hline

        \cite{2022AJ....164..249H} & Fid. GSE  & $6-60$ &
        $0.81_{-0.03}^{+0.03}$ & $0.73_{-0.02}^{+0.02}$ &
    $z[{24^\circ}_{-5^\circ}^{+6^\circ}] \rightarrow y'[{25^\circ}_{-3^\circ}^{+3^\circ}]$ &
        $1.7^{+0.16}_{-0.24}$\\

     \cite{2025arXiv251000095D} & Fid.   & $6-60$ & $0.88$ & $0.87_{-0.09}^{+0.05}$ & $z[{24^\circ}]\to y'[{43^\circ}_{-8^\circ}^{+22^\circ}]$ & $1.0$ \\
        
    \cite{Li:2025oxg} & Halo  & $8-200$ &
    $0.85^{+0.01}_{-0.01}$ & $0.74^{+0.01}_{-0.01}$ &
$z[{27^\circ}^{+1^\circ}_{-1^\circ}]\rightarrow y'[{44^\circ}^{+1^\circ}_{-1^\circ}]$ &
    $1.50^{+0.07}_{-0.07}$\\

     \cite{Nibauer:2025jvo} & Model 1 & $10-30$ &
$0.75_{-0.05}^{+0.05}$ & $0.70_{-0.04}^{+0.06}$ &
$z[{23^\circ}_{-7^\circ}^{+13^\circ}] \rightarrow y'[{18^\circ}_{-8^\circ}^{+5^\circ}]$ &
1.0 \\
      \hline

     \cite{Nibauer:2025jvo} & Model 2  & $10-30$ &
     $0.95_{-0.04}^{+0.06}$ & $0.65_{-0.05}^{+0.05}$ &
$z[{97^\circ}_{-8^\circ}^{+11^\circ}] \rightarrow y'[{56^\circ}_{-8^\circ}^{+8^\circ}]$ &
     1.0 \\
         
     \cite{Putney:2025mch} & Triaxial  & $4-12$ &
     $0.89_{-0.09}^{+0.09}$ & $0.33_{-0.06}^{+0.06}$ &
$z[{89^\circ}_{-2^\circ}^{+1^\circ}]\rightarrow y'[{52^\circ}_{-11^\circ}^{+11^\circ}]$ & $1.0$ \\
     \hline
\end{tabular}

 \label{tab:titlting}
 \end{table*}

\begin{figure*}
    \centering
    \includegraphics[width=0.98\linewidth]{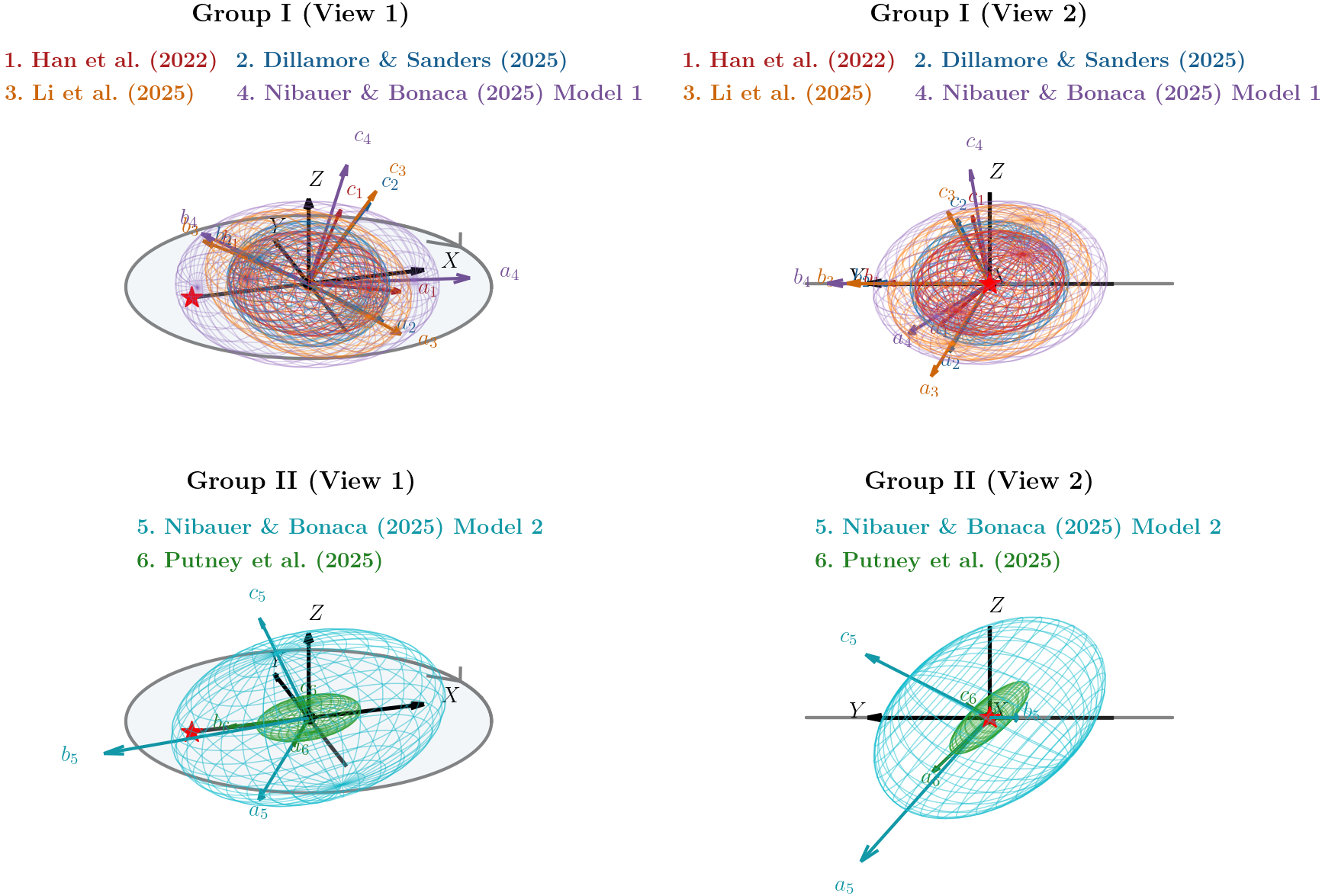}
    \caption{Tilted triaxial halos in~\tabref{titlting}. We show halos in Groups I and II in the upper and lower rows, respectively, with a 3D view (View 1, left column) and a view from the Sun to the GC (View 2, right column). For each halo, we set the elliptical radius to be  $r_{\min}$ of the model. The setup is the same as~\figref{benchmark}.}
    \label{fig:3D_halo}
\end{figure*}

In \tabref{titlting}, we show various dark matter or stellar halo models from previous studies. We include properties such as the range of the galactocentric distance, $r_{\min}$ and $r_{\max}$, of the data, the flattening parameters $p$ and $q$, rotation operations that act on an untilted configuration, and the (inner) cuspiness parameter $\gamma_{\rm inner}$. If a broken power law of the density profile is used~\cite{2022AJ....164..249H, Li:2025oxg}, we report the innermost density slope here, given that the innermost break radii are much larger than $r_\odot$.Figure~\ref{fig:3D_halo} shows the 3D configurations of all the halos provided in \tabref{titlting}, with $r_e = r_\text{min}$, at two views.

We interpret the rotation operations for each study according to the convention used in \appref{rotation}. We start from a configuration where the unit vectors of the triaxial halo's major, intermediate, and minor axes system coincide with the unit vectors of the galactocentric $X$, $Y$, $Z$ axes, respectively. We then perform a series of rotations of $A_1[\alpha_1] \to A_2 [\alpha_2]\to \cdots $, where $A_i$ represents the rotation axis and $\alpha_i$ represents the rotation angle with positive (negative) values for right-handed (left-handed) rotations. For active rotation, the principal axes often change their direction after each step of the rotation operation. We add a prime to the rotated axes to emphasize this change (e.g., $y'$).

The following list interprets and converts the rotation angles reported in earlier studies\footnote{We thank Adam Dillamore, Songting Li, Jacob Nibauer, and Eric Putney for discussions, in particular for clarifying the convention used in their studies.} into our convention described above.

\begin{enumerate}
       \item Reference~\cite{2022AJ....164..249H}, reports yaw and pitch angles of ${-24.33^\circ}_{-5.51^\circ}^{+4.94^\circ}$ and ${-25.39^\circ}_{-3.20^\circ}^{+3.11^\circ}$, respectively. Their values follow a passive rotation convention, which is the same as our convention. However, they define right-handed rotation as negative angles and left-handed rotation as positive angles, which is opposite to our sign convention. We therefore invert the signs of yaw and pitch angles, i.e., yaw of ${24.33^\circ}^{+5.51^\circ}_{+4.94^\circ}$ and pitch of ${25.39^\circ}^{+3.20^\circ}_{-3.11^\circ}$.

    \item Reference~\cite{Li:2025oxg}, reports yaw of ${-26.9^\circ}^{+1.0^\circ}_{-1.0^\circ}$ and pitch of ${-43.8^\circ}^{+0.7^\circ}_{-0.7^\circ}$. As in Ref.~\cite{2022AJ....164..249H}, these angles are given in the passive rotation convention, but the right-handed rotation is defined to be negative. To match our active convention, the yaw and pitch angle has been flipped to ${26.9^\circ}^{+1.0^\circ}_{-1.0^\circ}$ and ${43.8^\circ}^{+0.7^\circ}_{-0.7^\circ}$, respectively. 

        \item Reference~\cite{2025arXiv251000095D}, reports a yaw of ${204^\circ}$ and then a pitch of ${43^\circ}_{-8^\circ}^{+22^\circ}$ ($\beta_{dm}$). Their convention and ours are different in two aspects: (1) the definition of positive pitch (left-handed) is opposite to ours (right-handed), and (2) their yaw definition has a ${180^\circ}$ offset to ours. Given that we do not distinguish the ``head" and ``tail" of the triaxial halo, a yaw of ${204^\circ}$ is the same as a yaw of ${24^\circ}$, except that it will orient the $y'$-axis in the opposite direction, $-y'$. Because a left-handed pitch about $y'$ is equivalent to a right-handed pitch (the same pitch angle) about $-y'$, the translated pitch angle in our convention is still ${43^\circ}_{-8^\circ}^{+22^\circ}$.

\item Reference~\cite{Putney:2025mch}, reports a yaw of ${-89^\circ}_{-1^\circ}^{+2^\circ}$, adopting the same positive rotation angle convention as ours. However, their right-handed Galactocentric coordinate system places the Sun on the positive $X$-axis, whereas ours places it on the negative $X$-axis. This requires translating the yaw angle to ${89^\circ}_{-2^\circ}^{+1^\circ}$ to match our convention. No translation of the pitch angle is needed. 

\item Other studies, such as \cite{2023MNRAS.521.4936K}, parametrize the orientation of the tilted halo by specifying the direction of its principal axes in heliocentric coordinates $(\ell,b)$, rather than using yaw and pitch angles. We refer the reader to \cite{2025arXiv251000095D} for a comparison between the models of \cite{2023MNRAS.521.4936K} and those of \cite{2025arXiv251000095D} and \cite{Nibauer:2025jvo}.

\end{enumerate}

\section{ENERGY BINS AND THE POINT-SOURCE MASK}
\label{sec:masking}

We show the energy bin setup in~\tabref{mask}. For the point-source mask, we first classify point sources into two groups: those with a test statistic (TS) below 49 and those with a TS of 49 or greater. We mask each group with circular regions with radii $r_s$ and $r_l$, respectively. Both $r_s$ and $r_l$ depend on the photon energy, as shown in the last two columns of~\tabref{mask}.

\begin{table}[h]
     \centering
     \topcaption{The energy bin setup and the radii of the point-source masks,  $r_{s}$($r_{l}$), for point source with TS$<49$ (TS$\geq49$).}
     \label{tab:mask}
     
     \begin{tabular}{c |c|c c}
     \hline
        Energy Bin & $E_{\min}-E_{\max}$ [GeV] & $r_{s}$[\degree] & $r_{l}$[\degree] \\
     \hline
      0 & 0.275 -- 0.357 & 1.125 & 3.75 \\
      1 & 0.357 -- 0.464 & 0.975 & 3.25 \\ 
      2 & 0.464 -- 0.603 & 0.788 & 2.63 \\ 
      3 & 0.603 -- 0.784 & 0.600 & 2.00 \\
      4 & 0.784 -- 1.02 & 0.450 & 1.50 \\
      5 & 1.02 -- 1.32 & 0.375 & 1.25 \\
      6 & 1.32 -- 1.72 & 0.300 & 1.00 \\
      7 & 1.72 -- 2.24 & 0.225 & 0.750 \\
      8 & 2.24 -- 2.91 & 0.188 & 0.625 \\
      9 & 2.91 -- 3.78 & 0.162 & 0.540 \\
      10 & 3.78 -- 4.91 & 0.125 & 0.417 \\
      11 & 4.91 -- 10.8 & 0.100 & 0.333 \\
      12 & 10.8 -- 23.7 & 0.060 & 0.200 \\ 
      13 & 23.7 -- 51.9 & 0.053 & 0.175 \\
      
     \hline  
\end{tabular}     

\end{table}

\section{ALTERNATIVE ASSUMPTIONS OF TRIAXIAL AND ROTATED DARK MATTER PROFILE}
\label{app:AlternativeDMprofile_Nibauer_Bonaca}

\begin{figure}
    \centering
    \includegraphics[width=1.0\linewidth]{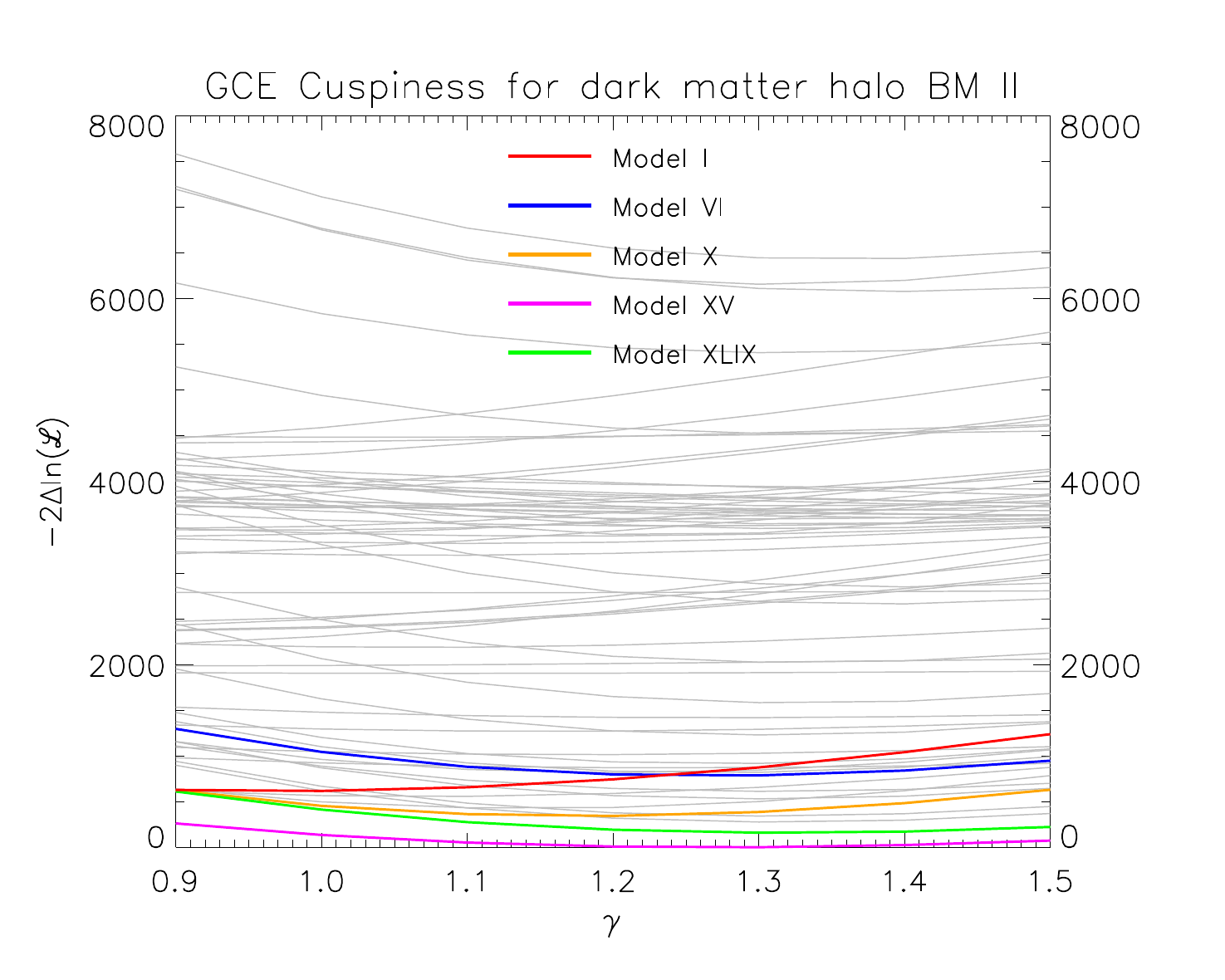}
    \vspace{-0.5cm}
    \caption{As in Fig.~\ref{fig:GCE_cuspiness} (top panel), the cuspiness $0.9 \leq \gamma \leq 1.5$, of the GCE morphology for an NFW-like triaxial dark matter annihilation profile that is rotated to our line of sight using the assumptions of Ref.~\cite{Nibauer:2025jvo} (i.e., BM II of Fig.~\ref{fig:example}). The best-fit models give a preference for $\gamma\simeq 1.2-1.3$.}
    \label{fig:GCE_cuspiness_NB}
\end{figure}

\begin{figure}
    \centering
    \includegraphics[width=1.0\linewidth]{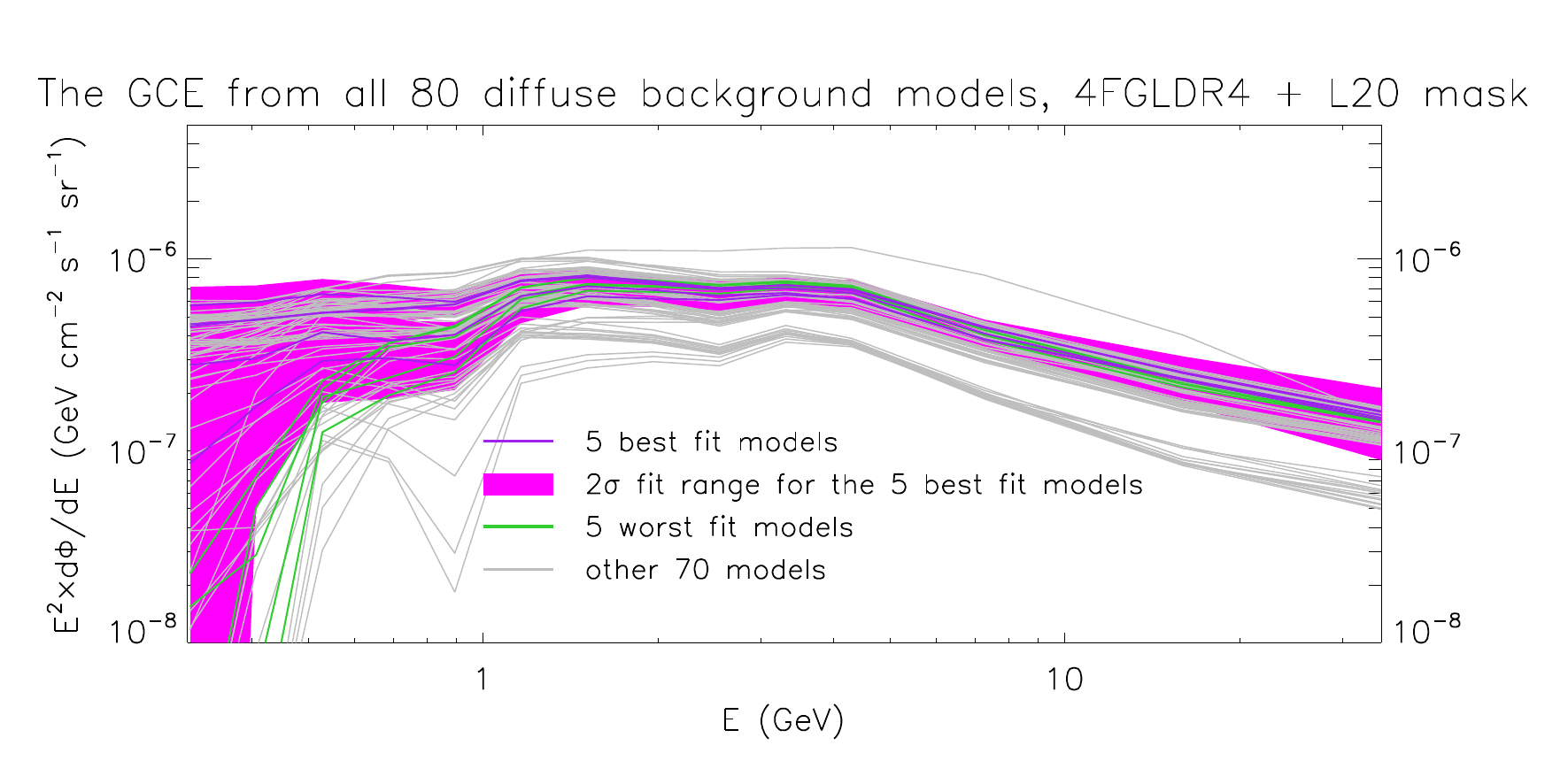}
    \caption{As with Fig.~\ref{fig:GCE_spectrum}, the GCE spectrum in the $40^{\circ}\times 40^{\circ}$ region of interest, assuming instead the triaxial and rotated dark matter profile with the assumptions of Ref.~\cite{Nibauer:2025jvo} (benchmark model II), with $\gamma=1.2$. Again, the GCE spectrum from the 80 diffuse emission models is effectively identical, choosing instead $\gamma=1.3$ or $\gamma=1.4$.}
    \label{fig:GCE_spectrum_NB}
\end{figure}

In this section, we present the results on the GCE cuspiness test and the GCE spectrum for all 80 GDE models, assuming that the GCE is due to the dark matter annihilation of BM II, based on model 2 of Ref.~\cite{Nibauer:2025jvo}. In Fig.~\ref{fig:GCE_cuspiness_NB}, we show the log-likelihood cuspiness test. 
As with BM I of Fig.~\ref{fig:GCE_cuspiness}, the best-fit results are for $\gamma \simeq 1.2-1.4$. 
In Fig.~\ref{fig:GCE_spectrum_NB}, we present the associated GCE spectrum from that BM II. The derived GCE spectrum is essentially the same for BM I and II, and in fact, also very similar to that derived assuming a spherical dark matter profile.  

\begin{figure}
    \centering
    \includegraphics[width=1.0\linewidth]{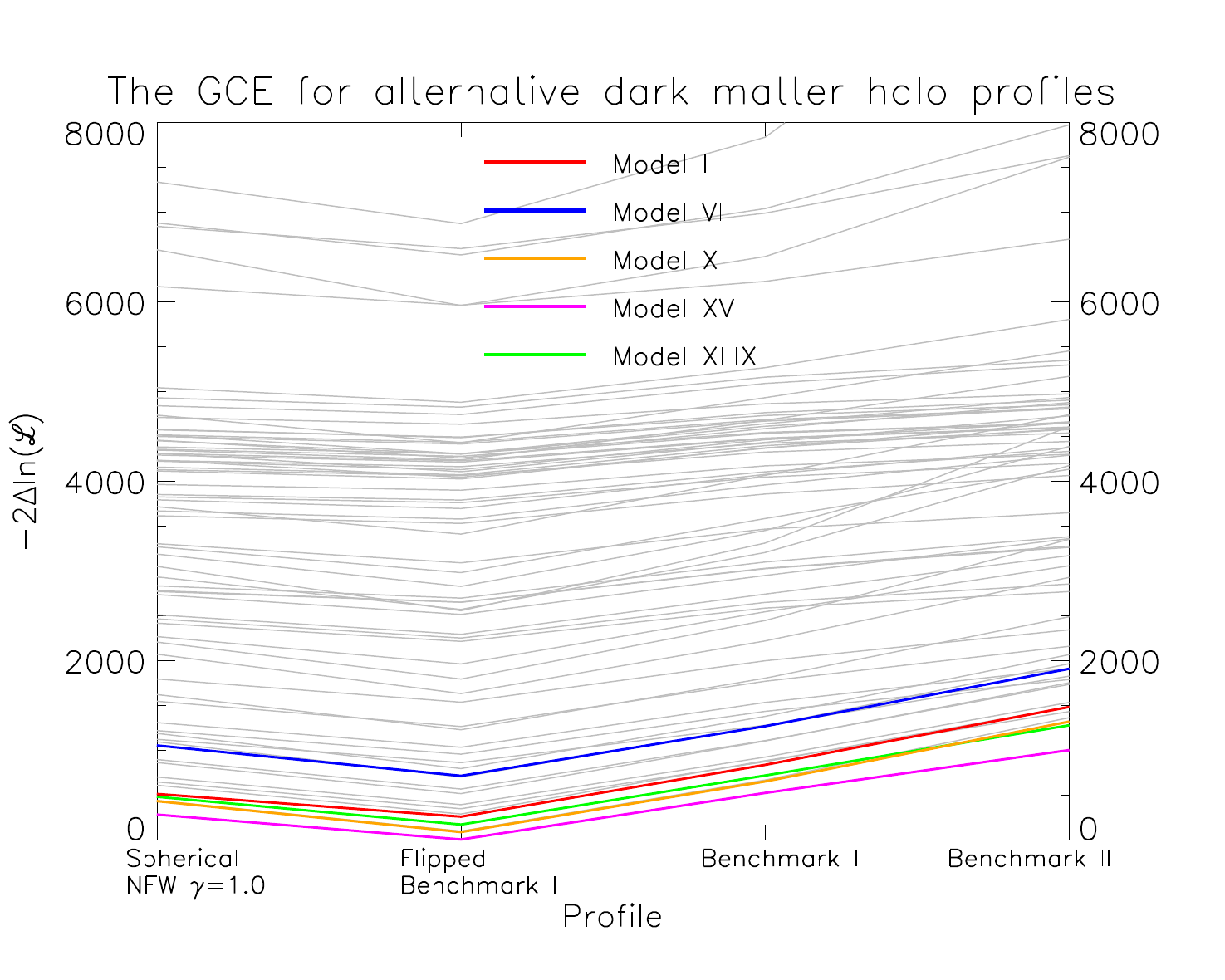}
    \vspace{-0.8cm}
    \caption{
    As in Fig.~\ref{fig:GCE_for_alternative_profiles} of the main text, we compare between alternative assumptions on the axes-ratios and rotations of the dark matter profile with $\gamma=1.0$.}
    \label{fig:GCE_for_alternative_profiles_gamma_1p3}
\end{figure}

In Fig.~\ref{fig:GCE_for_alternative_profiles_gamma_1p3}, we compare the alternative choices for the dark matter halo's triaxial and rotation assumptions for a cuspiness of $\gamma=1.0$ (in Fig.~\ref{fig:GCE_for_alternative_profiles} of the main text, we used $\gamma=1.2$). Like with the case of $\gamma=1.2$, there is a statistical preference for the flipped BM I, followed by the spherical dark matter profile, followed by BM I; with BM II assumptions being the least preferred. 
These results are independent of the GDE model used and are also confirmed when using instead $\gamma=1.3$ for all four sets of axes ratios and rotations.

\clearpage
\bibliography{reference}
\bibliographystyle{h-physrev.bst}

\end{document}